\documentstyle[aps,psfig,youngtab]{revtex}   
\newcommand{\half}[1]{\frac{#1}{2}} 
\newcommand{\prt}[1]{{(#1)}} 
 
\renewcommand{\>}{\rangle} 
\newcommand{\<}{\langle}

\newcommand{\fig}[1]{Fig.~\ref{#1}} 
\draft 
\begin{document} 
 
\title{Localization of eigenstates in a modified Tomonaga-Luttinger model} 
 
\author{Dimitry M. Gangardt and 
Shmuel Fishman}

\address{Department of Physics, Technion, 32000 Haifa, Israel} 
 
 
\maketitle 

\begin{abstract} 
We study the localization in the Hilbert space of a modified 
Tomonaga-Luttinger model. For the  standard version of this model, 
the states are found to be extended 
in the basis of  Slater determinants, 
representing the eigenstates of the non-interacting system. The 
linear dispersion which leads to the fact that these eigenstates 
are extended in the modified model is replaced by one with random 
level spacings modeling the complicated one-particle spectra of realistic models. 
The localization properties of the eigenstates are studied. The interactions 
are simplified and an effective one-dimensional Lloyd model is obtained. 
The effects of many-body  energy correlations are studied numerically. 
The eigenstates of the system are found to be localized in Fock space  for any 
strength of the interactions, but the localization is not exponential. 
\end{abstract} 

 
\section{Introduction} 
 
The concept of Anderson localization in the Fock space of 
many-body systems has been recently adopted for the studies of 
interacting electrons in finite-size conductors. The traditional 
field of localization consists of the investigation of electronic 
motion in disordered solids. These studies were initiated  by 
Anderson \cite{anders} who considered  tight-binding Hamiltonians 
on real-space lattices with random on-site energies. The behavior 
of the system was found to depend on the parameter $Z=W/V$, where 
$W$ is the typical variance of the random potential and $V$ is the 
hopping matrix element. For $Z$ above some critical value the 
eigenfunctions of the Hamiltonian are exponentially localized  with 
a characteristic scale $\xi$, called the localization length, 
which leads to the absence of diffusion of electrons. The Anderson 
problem inspired numerous studies (for reviews see 
\cite{leeram,thoul1,lifsh}). The scaling theory of localization \cite{aalr}
predicted the strong dependence of the localization properties on 
the dimension of the lattice. It was found, for instance, that in 
one spatial dimension the states are always localized for any 
strength of the hopping. 
 
The ideas of localization can be extended to the investigation of 
the properties  of interacting many-body Hamiltonians in the basis of 
eigenstates of the corresponding free system. The central problem 
considered there is the 
mixing of a particular state representing one particle excited 
above the Fermi surface with many-particle states responsible for 
emergence of a quasi-particle described by the usual Landau Fermi 
liquid theory. 
For this purpose the analogy between the 
one-particle and the many-body problems was demonstrated in 
\cite{altsh} by mapping the Hilbert space of interacting electrons 
onto an effective tight-binding model. Each site of the effective 
lattice represents the eigenstate of the non-interacting system 
and the bonds represent the interaction matrix elements. After 
several approximations, which consist mainly in neglecting the 
possibility of closed loops on this effective lattice, the 
resulting model was identified with the Cayley tree with branching 
number $K$ depending on the energy $\epsilon$ of the initial 
excitation, representing a particle excited above the Fermi sea. 
The Cayley tree  model was previously studied \cite{thoul2} and 
was found to undergo a localization transition for the value of 
parameter $Z=Z_c\approx K\ln K$. In the context of the original 
many-electron problem, the system is in the localized regime of 
the many-particle states on the states of the non-interacting 
system for energy below the lower threshold 
$\epsilon^{**}=\Delta\sqrt{g/\ln g}$, where $\Delta$ is the 
one-particle level-spacing and $g$ is dimensionless conductance, 
i.e. conductance in  units of $e^2/\hbar$. It was also argued that 
until the excitation energy $\epsilon$ reaches another critical 
value $\epsilon^*=\Delta\sqrt{g}>\epsilon^{**}$ the states are not 
completely delocalized, i.e. they do not mix large number of 
non-interacting states, but rather a small portion of them, 
corresponding to a subtree of the whole lattice. The last result 
was not confirmed in the subsequent studies \cite{mirl} of this 
model by super-symmetry methods. It was found there  that the 
states are  strongly correlated above the threshold 
$\epsilon>\epsilon^{**}$. 
 
In more recent numerical studies  of the interacting models 
\cite{berk,pascaud} it was observed that the choice of the type of 
the lattice affects  the localization-delocalization transition 
point. The crossover between the spectral statistics and the 
properties of eigenstates  such as inverse participation ratio 
were shown to be sensitive to the type of the lattice. It was 
doubted that the actual transition and the properties of states in 
the delocalized phase can be accounted for properly using the pure 
tree-like structure of the lattice. It was claimed in \cite{silv} that 
the infinite tree-like lattice used in \cite{altsh} may be  inadequate 
for description of the splitting of the quasi-particle peak into 
many-body constituents. The finiteness of the lattice used in this work 
results 
in the absence of the sharp transition from localized to 
delocalized states.  In the numerical work \cite{weid}, where the 
possibility of closed loops on the lattice was taken into account 
using the random-matrix theory 
 no delocalization transition was observed. Rather the 
participation ratio and spectral width of the eigenstates of the 
interacting system exhibited a smooth crossover from almost 
localized to delocalized states.

The exact solution for the delocalized phase (which corresponds to 
the strongly interacting regime) is not available in these studies 
in any limit . It would be useful to have a model which can 
clarify the nature of the extended states in the strongly 
interacting regime. The candidates for such a system are of course 
one-dimensional interacting system whose analytical solution is 
available for any strength of the interactions. The present work 
was inspired by the observation that the eigenstates of the 
simplest interacting problem, namely the chiral Tomonaga-Luttinger 
model \cite{haldane} are extended in the effective lattice of the 
non-interacting Slater determinants. This model is usually solved 
by the  bosonization technique, which can be viewed as a 
generalization of the solutions by Fourier transform of 
tight-binding models without disorder. This analogy can be traced 
more explicitly  if one considers the Hilbert space spanned by the 
Slater determinants as an effective lattice where each site is 
labeled by partition of some integer. Due to the conservation of 
total momentum the Hamiltonian of the chiral Tomonaga-Luttinger 
model is block-diagonal in the total momentum index $N$ and the 
linear dispersion implies that energies of all the states inside 
one block are degenerate and equal $N\Delta$. The interactions 
lift this degeneracy  mixing the Slater determinants into coherent 
superpositions which are described by the bosonic quantum numbers. 
Thus the bosonic eigenstates of the full interacting model can be 
viewed as a generalization of the extended Bloch states in the 
tight-binding models with degenerate on-site energies.

The linearization of the dispersion is the approximation which 
allows to solve the problem of interacting electrons in one 
dimension. The deviations from the linearity mixes the bosonic 
states, so they cannot any longer provide the solution to the 
model. In this work we  introduce the random dispersion into the 
Tomonaga-Luttinger model and study the properties of its 
eigenstates. It is expected to model a system where the single 
particle spectrum is complicated. The question we address is 
whether this model exhibits localization, since the randomness 
introduced into the dispersion leads to  randomness of the diagonal 
matrix elements of the Hamiltonian analogous to the diagonal 
disorder introduced into tight-binding Anderson models. We have 
found that the states remain localized in the basis of Slater 
determinants, i.e. resemble  the states of the non-interacting 
systems. 
 
The outline of this paper is as follows. In section 2 we review some 
known facts about the standard Tomonaga-Luttinger model. We 
describe the structure of the Hilbert space as an effective 
lattice and discuss the bosonization solution in this context.  In 
section 3 we introduce randomness into the dispersion law of the 
Tomonaga-Luttinger model and simplify the interaction matrix. Then 
we solve the resulting model with assumption of independence of 
the on-site energies. Section 4 is devoted to the study of 
relevance of the correlations between the energies. In the last 
section the results are discussed and further investigations are 
proposed.

\section{Hilbert space of Tomonaga-Luttinger model} 
 
In this section we review a model exactly solvable for any 
interaction strength. The simplest model of interacting fermions 
is the one-branch (chiral) Tomonaga-Luttinger model 
\cite{haldane}. Let $n$ label the one-particle orbital with 
convention that in the ground state $|0\>$ of the non-interacting 
model, $n=1$ denotes the first unoccupied state. Since a very large 
number of states is occupied (the number of fermions is large) 
the number $n$ can be allowed to run from $-\infty$ to $+\infty$. 
Another consequence is that the one-particle spectrum is assumed 
to be equidistant: $E_n=n \Delta $. Both the linear dispersion and 
the fact that  the spectrum is unbound are crucial for the 
solubility of the  model. Introducing creation and annihilation 
operators $c^\dagger_n$, $c_n$ for a fermion on the orbital $n$, 
with usual anti-commutation relations $\{c_n, 
c^\dagger_{n'}\}=\delta_{n,n'}$ (other anti-commutators vanish) 
the Hamiltonian of the problem can be written as 
\begin{equation} 
\hat{H}=\hat{H}_0+\hat{U} 
=E_0 + \sum_{n=-\infty}^{+\infty} E_n\, :c^\dagger_n c_n: + 
\sum_{m=1}^{+\infty}\sum_{n,n'=-\infty}^{+\infty} 
 V_m  c^\dagger_{n+m} c_n c^\dagger_{n'-m} c_{n'} 
\label{Ham} 
\end{equation} 
The normal ordering of operators having infinite expectation values is 
defined as $:\hat{O}: = \hat{O}-\<0|\hat{O}|0\>$. As a result 
the constant $E_0$ absorbs the (infinite) free ground state energy together 
with the Hartree term: 
\begin{equation} 
E_0=\<0|\hat{H}|0\> = \sum_{n\leq 0}E_n \<0|c^\dagger_n c_n|0\>+ 
\half{V_0}  \sum_{n,n'\leq 0}\<0|c^\dagger_n c_n c^\dagger_{n'} c_{n'}|0\> 
\label{e0} 
\end{equation} 
The operator of ``total momentum'' 
\begin{equation} 
\hat{P}=\sum_{n=-\infty}^{+\infty} n \, :c^\dagger_n c_n: 
\label{Mom} 
\end{equation} 
commutes with the full Hamiltonian. Due to the linear dispersion 
this operator is proportional to the free part of the Hamiltonian, i.e. 
the first term on the RHS of (\ref{Ham}). 
 
\begin{figure} 
\centerline{\psfig{figure=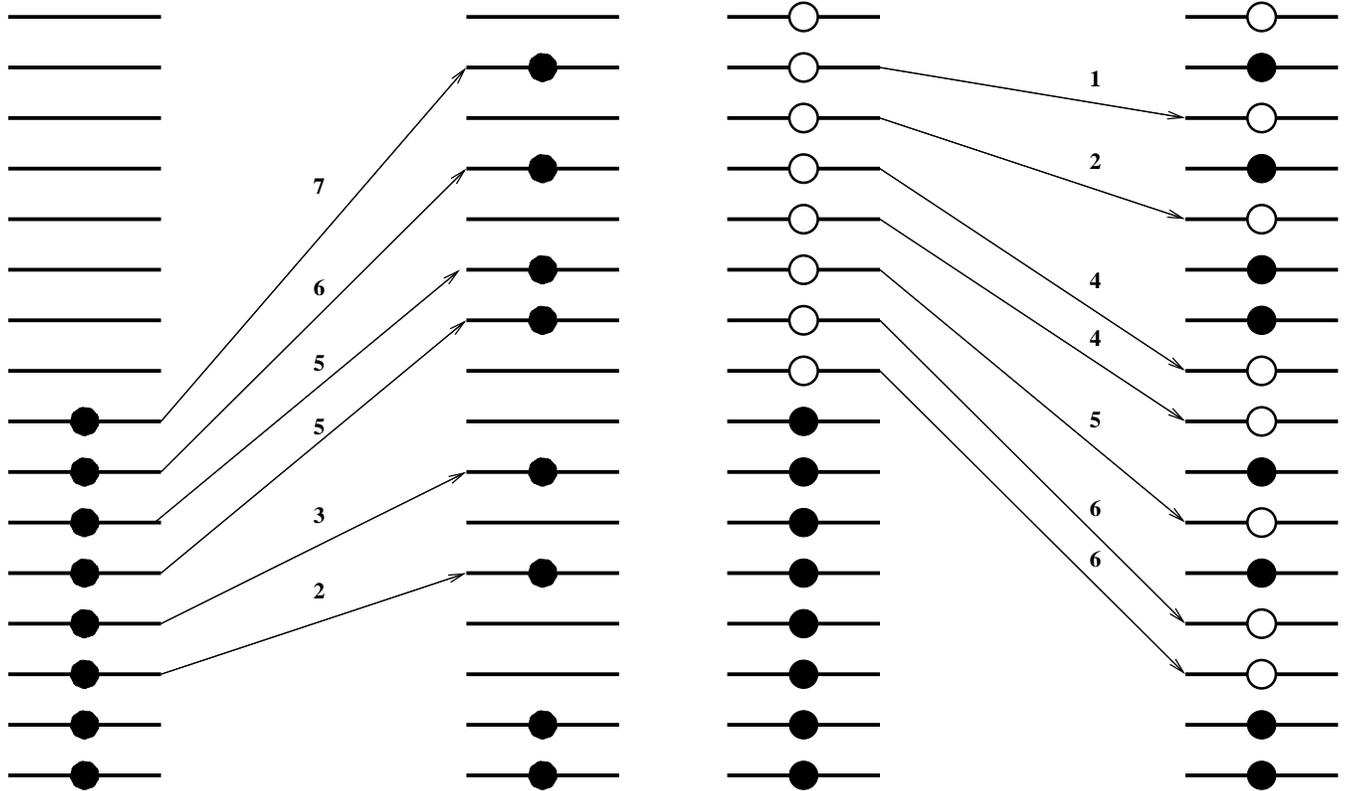,width=17.8cm}} 
\hspace{10pt}
\caption{Correspondence between Slater determinant and partition 
$\prt{\lambda}=\prt{7,6,5,5,3,2}$. The dual partition 
$\tilde{\prt{\lambda}}=\prt{6,6,5,4,4,2,1}$ represents shifts of the holes.} 
\label{shifts} 
\end{figure}

Instead of solving this model  by  the standard method  of 
bosonization using identities among second quantized operators 
(see for example \cite{haldane}) we prefer to reformulate our 
interacting problem in order to relate it to the problems usually 
discussed within the framework of Anderson localization. We start 
by labeling the eigenstates of non-interacting problem, which are 
the familiar Slater determinants, by shifts of particles with respect 
to the non-interacting ground state as shown 
in the \fig{shifts}: the uppermost particle is shifted up 
$\lambda_1$ levels, the second one is shifted up $\lambda_2$ levels etc. 
The numbers $\lambda_i$ are defined to satisfy $\lambda_i\ge\lambda_{i+1}$ for each $i$. 
If  a particle is not shifted it is not recorded. 
The set $\prt{\lambda}$ of positive integers 
satisfying $\sum \lambda_i=N$ forms the set of partitions of 
$N$ and is conveniently displayed by a Young diagram 
--- array with $\lambda_i$ boxes in the $i$-th row. Following the standard notations 
in the mathematical literature the numbers 
$\lambda_i$ will be  referred to as \emph{parts} in what follows. 
For example, the partition of $N=28$, $\prt{\lambda}=\prt{7,6,5,5,3,2}$ 
sometimes written as $\prt{7,6,5^2,3,2}$ that is displayed in \fig{shifts} is 
represented graphically as 
\begin{displaymath} 
\prt{\lambda}\rightarrow\yng(7,6,5,5,3,2) 
\end{displaymath} 
This pictorial representation of Slater determinants has nice properties 
under particle-hole interchange: recording shifts of the  holes 
(which are moved downwards in energy) rather than particles in 
\fig{shifts} yields the partition 
$\tilde{\prt{\lambda}}=\prt{6,6,5,4,4,2,1}$ whose graphical representation is 
just the transpose of the Young diagram for $\prt{\lambda}$: 
\begin{displaymath} 
\tilde{\prt{\lambda}}\rightarrow\yng(6,6,5,4,4,2,1) 
\end{displaymath} 
Partition $\tilde{\prt{\lambda}}$ is said to be conjugate or dual to 
$\prt{\lambda}$. This duality results from the fact that each move of a particle 
upwards is accompanied by a move of a hole downwards. 
 
The eigenstates of the non-interacting Hamiltonian labeled by 
partitions of $N$ are degenerate  and their energy is given by 
\begin{equation}\label{Epart} 
\sum_i \left(E_{1-i+\lambda_i}-E_{1-i}\right)= \Delta \sum_i 
\lambda_i=N\Delta 
\end{equation} 
The interactions conserve this number. The Hamiltonian 
(\ref{Ham}) is therefore block-diagonal and does not mix the 
partitions of  different $N$ so we can restrict ourselves to the 
subspace of the Hilbert space spanned by partitions of $N$.

The dimensionality of this subspace is given by 
$p(N)$ --- the number of partitions of $N$. There is no closed analytical 
expression for this number but the Euler expression \cite{andrews} 
for the generating function 
\begin{equation} 
p(q)=\sum_{N=1}^{\infty} p(N)\, q^N = \prod_{l=1}^{\infty} \frac{1}{1-q^l} 
\label{Gen} 
\end{equation} 
allows to obtain the asymptotic behavior of $p(N)$ for large $N$ and it is 
given by the formula of Hardy-Ramunajan: 
\begin{equation} 
p(N)\sim \frac{e^{\pi\sqrt{2N/3}}}{4\sqrt{3}N} 
\label{HR} 
\end{equation} 
It demonstrates that the number of ``sites'' in the subspace increases rapidly 
with $N$. 
 
 Since for a given subspace $N$ the free part of the Hamiltonian is 
proportional to the unit matrix the solution consists in 
diagonalization of the interacting part $U$. The interaction term 
in (\ref{Ham}) is conveniently represented in our picture as 
effective binding between the sites. In order to present the 
results in a compact and transparent way we drop the requirement 
on a partition labeling the basis state to be represented by a 
non-increasing sequence. Instead we extend the notion to any 
finite sequence of integers (either positive or negative or zero) 
and use the following rules to relate this sequence to the 
standard (non-increasing) partition: 
\begin{enumerate} 
\item {\em In any sequence $\prt{\lambda_1,\lambda_2,\ldots}$ 
two consecutive parts may be interchanged provided that the preceding part 
is decreased by unity and the succeeding part increased by unity, the state 
vector which corresponds to this partition acquires a minus sign}, i.e. 
\begin{equation} 
|\prt{\lambda_1,\ldots,\lambda_i,\lambda_{i+1},\ldots}\>=- 
|\prt{\lambda_1,\ldots,\lambda_{i+1}-1,\lambda_i+1,\ldots}\> 
\end{equation} 
 
\item {\em If any part exceeds by unity the preceding part the partition corresponds to zero 
vector}, i.e. 
\begin{equation} 
|\prt{\lambda_1,\ldots,\lambda_i,\lambda_i+1,\ldots}\>=0 
\end{equation} 
 
\item {\em If the last part of the sequence is negative the partition corresponds to the zero 
vector}. 
 
\end{enumerate} 
 
These rules, used in the theory of symmetric functions 
\cite{littlewood} are in fact nothing but a standard properties of 
Slater determinants under action of second quantized operators. 
Rule 1 results from the fact that $i$-th particle is excited to 
the level $n_i$ such that 
\begin{displaymath} 
n_i = \lambda_i+1-i 
\end{displaymath} 
with the energy $E_i= n_i\Delta$ 
while the $(i+1)$-th particle is excited to the level 
\begin{displaymath} 
n_{i+1}= \lambda_{i+1}-i 
\end{displaymath} 
and interchanging consecutive rows in a (Slater) determinant results 
in a sign change. Rule 2 results from the fact that 
$\lambda_{i+1}=\lambda_i+1$ implies $n_{i+1}=n_i$ which 
contradicts the Pauli principle. Rule 3 reflects the fact that 
negative $\lambda_i$ describe (forbidden) transitions to occupied 
states inside the Fermi sea. 
 
\begin{figure} 
\centerline{\psfig{figure=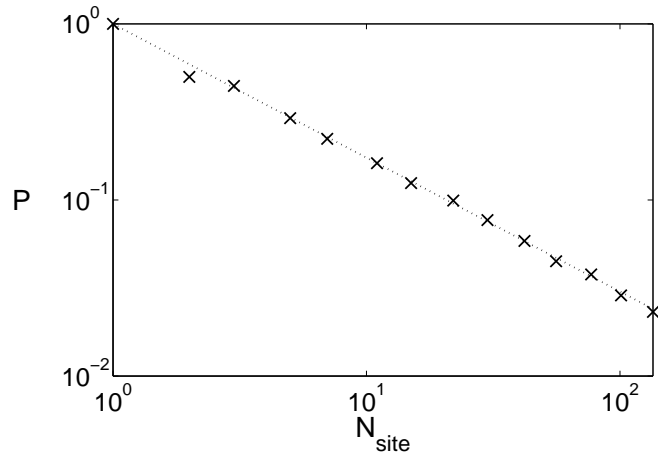,width=8.6cm}}
\hspace{10pt} 
\caption{The inverse participation ratio (\ref{invpart}) as function 
of number of sites $N_{site}=p(N)$. The solid line represents the  fit 
$P=1/N_{site}^\alpha$, where $\alpha\approx 0.8$} 
\label{invp} 
\end{figure}

Let us denote by $\eta_\prt{\lambda}$ the factors $-1,0$ or $+1$ 
acquired  when bringing a partition $\prt{\lambda}$ to its standard 
form. In these notations the matrix element of the 
interactions between the partitions $\prt{\lambda}$ and $\prt{\mu}$ can be 
represented as 
\begin{equation} 
U_{\prt{\lambda},\prt{\mu}} = \<\prt{\lambda}|\hat{U}|\prt{\mu}\> = 
\sum_m V_m \sum_{i,j} \<\prt{\ldots\lambda_i-m\ldots}| 
\prt{\ldots\mu_j-m\ldots}\>. 
\label{Int1} 
\end{equation} 
Since the basis of Slater determinants is orthonormal the following expression 
is obtained 
\begin{equation} 
U_{\prt{\lambda},\prt{\mu}} = \sum_m V_m \sum_{i,j} 
\eta_\prt{\ldots\lambda_i-m\ldots}\eta_\prt{\ldots\mu_j-m\ldots} 
\delta_{\prt{\ldots\lambda_i-m\ldots}, \prt{\ldots\mu_j-m\ldots}}. 
\label{Int2} 
\end{equation} 
The eigenstates of $U$ are 
\begin{equation}\label{lstate} 
  |\prt{l}\>=\sum_\prt{\lambda}\chi^\prt{\lambda}_\prt{l}|\prt{\lambda}\> 
\end{equation} 
where $\chi^\prt{\lambda}_\prt{l}$ are the characters of the 
irreducible representation $\prt{\lambda}$ of an element of the 
symmetric group $S_N$ belonging to the class $\prt{l}$. 
 For this purpose it is shown in Appendix A  that 
\begin{eqnarray} 
\sum_{i,j}\eta_\prt{\ldots\lambda_i-m\ldots} 
\eta_\prt{\ldots\lambda_i-m\ldots\lambda_j+m\ldots} 
\chi^\prt{\ldots\lambda_i-m\ldots\lambda_j+m\ldots}_\prt{l}= 
 mn_m \chi^\prt{\lambda}_\prt{l} 
\label{theproperty} 
\end{eqnarray} 
where $n_m$ is  the 
number of times the cycle $m$ appears in the  class $\prt{l}$. 
This leads to 
\begin{equation}\label{U} 
  \hat{U}|\prt{l}\> = \sum_m V_m m n_m 
  \sum_\prt{\lambda}\chi^\prt{\lambda}_\prt{l}|\prt{\lambda}\>= 
  \left(\sum_m V_m m n_m\right) |\prt{l}\>. 
\end{equation} 
The state $|\prt{l}\>$ is therefore an eigenstate of the Hamiltonian 
(\ref{Ham}) with the eigenvalue 
\begin{equation}\label{ev} 
  E_\prt{l} = E_0+N\Delta+\sum_m V_m m n_m . 
\end{equation} 
The orthogonality of the characters implies that the states 
$|\prt{l}\>$ form an orthogonal basis, namely: 
\begin{equation}\label{orthog} 
  \<\prt{l}|\prt{l'}\>=\frac{g}{g_\prt{l}}\;\delta_{\prt{l},\prt{l'}} 
\end{equation} 
where $g=N!$ is the total number of elements in $S_N$ and the 
number of elements in the class $\prt{l}$ is 
\begin{equation} 
g_\prt{l} = N!\prod_{m=1}^\infty \frac{1}{m^{n_m} n_m!}. 
\label{gl} 
\end{equation} 
We normalize the states $|\prt{l}\>$ as 
\begin{equation}\label{norml} 
  |\prt{l}\>=\sqrt{\frac{g_\prt{l}}{g}} 
  \sum_\prt{\lambda}\chi^\prt{\lambda}_\prt{l}|\prt{\lambda}\> 
\end{equation} 
so that they form an orthonormal basis. These are standard bosonic 
states usually labeled by the occupation numbers $n_m$ that are 
the eigenstates of the interacting Hamiltonian (\ref{Ham}). The 
transformation with the matrix 
\begin{equation} 
\<\prt{\lambda}|\prt{l}\>=\sqrt{\frac{g_\prt{l}}{g}}\chi^\prt{\lambda}_\prt{l} 
\label{amplitude} 
\end{equation} 
is the unitary transformation of the basis vectors in our Hilbert 
space, which diagonalizes the interacting Hamiltonian (\ref{Ham}). 
When the free part of the Hamiltonian is degenerate in the 
subspace $N$ the states $|\prt{l}\>$ play the role of the extended 
states (in the Hilbert space). In order to verify this statement 
we calculated numerically the inverse participation ratio $P$, 
defined as 
\begin{equation} 
P=\sum_\prt{\lambda} |\<\prt{\lambda}|\prt{l}\>|^4 
\label{invpart} 
\end{equation} 
for some arbitrary states $\prt{l}$.  Although no closed expression is 
available for the general amplitude (\ref{amplitude}) of the state 
$|\prt{l}$ on the site $\prt{\lambda}$, the beautiful recurrent 
relation of the characters of the symmetric group 
(\ref{recurrence}) (based on the Frobenius formula) was 
implemented numerically to calculate these amplitudes. 
The inverse participation ratio averaged over the bosonic states $\prt{l}$ 
is  shown in \fig{invp} for different values of 
$N$. 
As the number of sites $N_{site}=p(N)$ increases the inverse participation ratio 
$P$ was found to 
decay  as $P=1/N_{site}^\alpha$, where $\alpha\approx 0.8$. This allows us to 
conclude that the bosonic states $\prt{l}$ are indeed extended in the basis of free 
Slater determinants $\prt{\lambda}$.

\section{Simplified Interactions} 
 
The existence of extended states in the chiral Tomonaga-Luttinger 
model can be attributed to the degeneracy of the diagonal elements 
of the Hamiltonian (\ref{Ham}), since the interactions 
(\ref{Int2}) have short-range in the space of partitions.  In 
order to define more precisely what we mean by short-range in our 
Hilbert space we have to order the partitions in some order. Then 
the interaction can be shown to connect only  blocks of elements 
that are not too far one from the other. 
 
It is  natural  to order the partitions in the lexicographic 
order, i.e. if $\lambda_i=1,2,3\ldots$ numbers the letters in an 
alphabet the lexicographic order of partitions corresponds to the 
order of words in Arabic or Hebrew dictionary. For example, the 11 
partitions of $N=6$ in the lexicographic order are: 
\begin{equation} 
\begin{array}{cr} 
 k=1 & (1,1,1,1,1,1) \\ 
  & \\ 
 k=2 & (2,1,1,1,1) \\ 
  & \\ 
 k=3 & (3,1,1,1) \\ 
  & \\ 
 k=4 & (2,2,1,1) \\ 
     & (4,1,1) \\ 
  & \\ 
 k=5 & (3,2,1) \\ 
     & (5,1) \\ 
  & \\ 
 k=6 & (2,2,2) \\ 
     &(4,2) \\ 
     &(3,3) \\ 
     &(6) 
\end{array} 
\label{Partitions} 
\end{equation} 
The partitions are grouped by $k$---the number of 1's (that appear 
always on the right). The $k$-th group contains partitions of the 
form 
\begin{equation}\label{groupk} 
  \prt{\lambda_1,\lambda_2,\ldots\lambda_l,1^{N-k}} 
\end{equation} 
with constraint $\lambda_i\geq2$, with the only exception for the 
first group containing $(1^N)$. The typical Young diagram of a 
partition in the $k$-th group is of the form 
\begin{eqnarray} 
\Yvcentermath1
\Yinterspace0ex plus 0ex
\nonumber &&\hspace{-1.7em}\yng(6,4,4,3)\\ 
 N-k \left\{\begin{array}{l}\vdots \\
\yng(1,1,1,1)  
\end{array}\right. 
\label{example} 
\end{eqnarray} 
To calculate the number of partitions in the $k$-th group we note 
that exactly $p(k)$ partitions have at least $N-k$ ones. Among 
them $p(k-1)$ have at least $N-(k-1)$ ones and the rest 
$n(k)=p(k)-p(k-1)$ have \emph{exactly} $N-k$ ones. Therefore the 
$k$-th group contains $n(k)$ partitions.

\begin{figure} 
\centerline{\psfig{figure=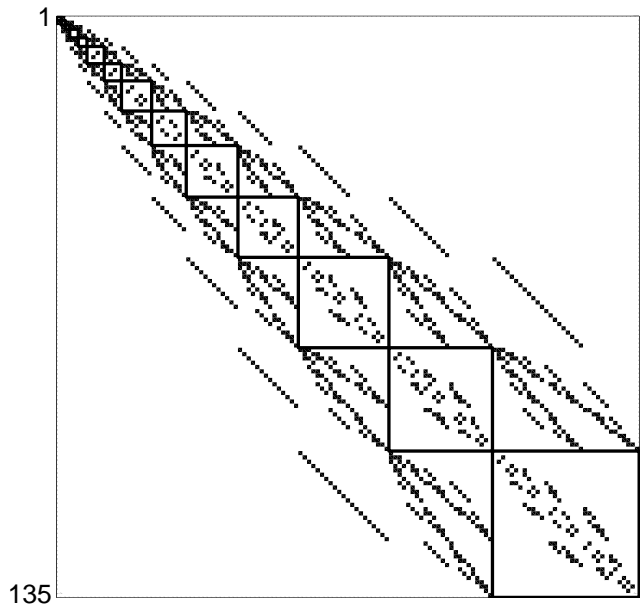,width=8.6cm}}
\hspace{10pt} 
\caption{The 
matrix of the interaction (\ref{Int3}). The blocks corresponding 
to different groups of partitions are shown. } \label{matrix} 
\end{figure}

Usually the matrix element of the interaction potential $V_m$ is 
a decreasing function of the momentum transfer $m$, with some 
effective range $m_c$. In what follows it will be convenient to 
analyze a model where the only non-zero matrix element corresponds 
to $m=1$ with $V_1=V$. It will not affect qualitatively our results, 
since we are primarily interested in the limit of large $N$ 
subspace. The matrix element of this interaction is given by the 
simplified version of (\ref{Int2}) 
\begin{equation} 
U_{\prt{\lambda},\prt{\mu}} = 
 V\sum_{i,j} \delta_{\prt{\ldots\lambda_i-1\ldots}, 
\prt{\ldots\mu_j-1\ldots}}. 
\label{Int3} 
\end{equation} 
The  factors $\eta$ are omitted since the partitions 
$\prt{\ldots\lambda_i-1\ldots}$ and $\prt{\ldots\mu_j-1\ldots}$ are either already in 
the standard (non-increasing) form or corresponds to zero vector 
and do not contribute to the sum. 
Therefore the interaction couples partitions $\prt{\lambda}$ and 
$\prt{\mu}$ that differ in two of the $\lambda_i$'s, namely 
$\lambda_i-\mu_i=1$ and $\lambda_j-\mu_j=-1$. The Young diagram of 
the partition $\prt{\lambda}$ is obtained from the one of the 
partition $\prt{\mu}$ by moving one square under the constraint 
that the new Young diagram is legitimate. Consider the $k$-th 
group. If $\mu_i>1$ and $\mu_j>2$ the coupling is between 
partitions inside this group. This corresponds to moving a square 
between $\mu_i>1$ rows  of the Young diagram. In order to couple 
to neighboring groups a square corresponding to $\mu_i=1$ should 
be moved. Moving a square from/to any row with $\mu_j>1$ to/from 
the region where $\mu_i=1$ leads to a coupling to the neighboring 
group $k\pm 1$. The exception is if a square is moved from 
$\mu_i=2$ (that has only two squares and after the move only one 
is left) and is attached as the last square in the tail of 
$\mu_j=1$, or if the last square is moved  to  $\mu_j=1$ turning 
it into a $\mu_j=2$ row. In this case the length of the tail of 
$\mu_j=1$ is changed by two, i.e. the coupling is to the group 
$k\pm 2$. These are the only moves of one square, therefore only 
the groups $k'$ such that $|k-k'|\leq 2$ are coupled to the $k$-th group and the 
interaction between the various blocks of partitions is of short 
range. \fig{matrix} shows the structure of the interaction matrix 
for $N=14$.

It is therefore natural to expect that the diagonal disorder will 
destroy the extended states of the chiral Tomonaga-Luttinger 
model. In the original model in the absence of interactions the 
eigenenergies $E_\prt{\lambda}$ are independent of the partition 
and take the constant value $E_\prt{\lambda}=E_0+N\Delta$. We 
introduce a modification and assume that the $E_\prt{\lambda}$ are 
independent random variables. This assumption is made for the sake 
of simplicity. It cannot be correct since there are $p(N)$ values 
of $E_\prt{\lambda}$ that depend on $2N$ level spacings. An 
assumption that is more physical will be introduced in the 
following section.

In order to be able to solve the model with random energies we 
simplify further the interaction (\ref{Int3}). From the experience 
in research of Anderson localization we expect that the behavior 
of the eigenvectors will not be affected by the precise form of 
the short-range off-diagonal matrix elements, so we make the 
following simplifications. We assume that the effect of 
transitions inside the same group of partitions is to delocalize 
the states within the group. The localization within a group is 
expected only to enhance the overall localization. Next, we 
neglect the coupling to the $k+2$-th group. The reason is that for 
a given partition $\prt{\lambda}$ there exists only one coupling 
term  of this kind. It couples 
\begin{equation} 
\left((\ldots),1^{N-k}\right)\;\;\mbox{and}\;\; 
\left((\ldots)',2,1^{N-k-2}\right) \label{Trans1} 
\end{equation} 
This should be compared it with the couplings of $\prt{\lambda}$ 
to the next group: 
\begin{equation} 
\left((\ldots),1^{N-k}\right)\rightarrow 
\left((\ldots)',1^{N-k-1}\right) 
\label{Trans2} 
\end{equation} 
The number of such couplings is given by the number of different 
$\lambda_i\ge 2$ in  the partition $\prt{\lambda}$, which is 
generally much larger than 1 for large enough $k$. It is equal to 
the number of rows with $\lambda_i\geq 2$ in the Young diagram of 
the partition. Consider for example the partition 
$\prt{6,4^2,3,1^4}$ of $N=21$ belonging to he  the 17-th group  corresponding to 
the Young diagram (\ref{example}). It has (apart of ones) three 
different parts: 6,4,3 and is coupled to 3 partitions in the 18-th 
group: 
\begin{equation} 
\begin{array}{lll} 
  \yng(7,4,4,3,1,1,1) & \yng(6,5,4,3,1,1,1) & \yng(6,4,4,4,1,1,1) \\ 
  & & \\ 
   \prt{7,4^2,3,1^3} & \prt{6,5,4,3,1^3} & \prt{6,4^3,1^3} 
\end{array} 
\label{couplings} 
\end{equation} 
Another way to visualize the number of coupling is to count the 
number of concave corners in the Young diagram that is equal to 
the number of parts namely lines of different length larger than unity. 
 
In  Appendix B we show that $d(k)$ --- the  number of different 
parts in all partitions of $k$ with $\lambda_i\ge 2$ is exactly 
$p(k-2)$ 
--- the number of (any) partitions of $(k-2)$. Therefore the mean 
number of transitions from  the $k$-th group to the $k+1$-th group 
is given by $t(k)=d(k)/n(k)=p(k-2)/\left(p(k)-p(k-1)\right)$. 
 
The Schr\"odinger equation for our simplified model can be written 
as a tight-binding form  where the groups of partitions are the 
effective sites: 
\begin{equation} 
(E-E_\prt{\lambda})\psi\prt{\lambda}_k = \sum_{\prt{\mu}} 
U_{\prt{\lambda},\prt{\mu}} \psi\prt{\mu}_{k-1} + \sum_{\prt{\nu}} 
U_{\prt{\lambda},\prt{\nu}} \psi\prt{\nu}_{k+1} 
\label{Schr1} 
\end{equation} 
where $\psi\prt{\lambda}_k$ denotes the value of the wave function 
on site$\prt{\lambda}$ in the $k$-th group. Let us assume that the 
wave function is extended within the $k$th group. We introduce a 
simpler model that is expected to exhibit similar localization 
properties \cite{pichard}. We replace the interaction 
$U_{\prt{\lambda},\prt{\mu}}$ by an average interaction that 
couples each partition in the $k$-th group to all partitions in 
the group $k\pm 1$, 
\begin{equation} 
V_k=\frac{\sum_\prt{\nu} U_{\prt{\lambda},\prt{\nu}}}{n(k+1)} 
\label{effInt1} 
\end{equation} 
In this approximation (\ref{Schr1}) reduces to 
\begin{equation} 
(E-E_\prt{\lambda})\psi\prt{\lambda}_k =  V_{k-1} \sum_{\prt{\mu}} 
\psi\prt{\mu}_{k-1} + 
 V_{k} \sum_{\prt{\nu}}  \psi\prt{\nu}_{k+1} 
\label{Schr2} 
\end{equation} 
The numerator in (\ref{effInt1}) can be estimated as $V$ times the 
mean number of couplings with  partitions in $(k+1)$-th group: 
\begin{equation} 
V_k=V \frac{t(k)}{n(k+1)}=V \frac{d(k)}{n(k)n(k+1)} 
\label{effInt2} 
\end{equation} 
 
Denoting the sum of the wave-functions over a group by 
$C_k=\sum_{\prt{\lambda}}  \psi\prt{\lambda}_k$ we find with the help of 
(\ref{Schr2}) that 
\begin{equation} 
{\cal E}_k C_k = V_{k-1} C_{k-1}+V_k C_{k+1} 
\label{effSchr} 
\end{equation} 
where 
\begin{equation} \frac{1}{{\cal E}_k} = \sum_\prt{\lambda} 
\frac{1}{E-E_\prt{\lambda}} \label{Ek} 
\end{equation} 
and the sum is over all partitions in the group $k$.

For random independent $E_\prt{\lambda}$ all the ${\cal E}_k$ are independent random 
variables and the tight-binding model 
(\ref{effSchr}) is an Anderson model at zero energy. In the spirit 
of transfer matrix method it is expected that  also for ${\cal E}_k$ 
given by (\ref{Ek}) similar localization properties will be found. 
The properties of the  solution of this one-dimensional model can 
be analyzed within the framework of Anderson localization in 
one-dimensional systems \cite{ishii}. 
 
One can take the uniform distribution of energies that turns out 
to be convenient for calculations. Remembering that each many-body 
energy is a sum of $N$ level spacings we assume the following 
distribution: 
\begin{equation} 
P(E)=\left\{ \begin{array}{cc} 
1/\sqrt{N}\Delta, & -\sqrt{N}\Delta/2<E<\sqrt{N}\Delta/2 \\ 
0, & \mbox{otherwise} \end{array}\right. 
\label{PE} 
\end{equation} 
 
 For a variety of distributions 
the distribution of ${\cal E}_k$ tends to the Cauchy distribution: 
\begin{equation} 
P({\cal E}_k)=\frac{1}{\pi}\frac{\delta_k(E)}{{\cal E}_k^2 + \delta_k(E)^2} 
\label{Cauchy} 
\end{equation} 
where $\delta_k(E)=\delta(E)/n(k)$ and for (\ref{PE}) $\delta(E)=\sqrt{N}\Delta/\pi$. 
 
Multiplying  the equation (\ref{effSchr}) by $n(k)n(k+1)/(V d(k))$ 
and observing that in the limit $k\rightarrow\infty$ one finds 
$n(k)/n(k+1)\rightarrow 1$ and $d(k)/d(k-1)\rightarrow 1$ so 
that (\ref{effSchr}) takes the form 
\begin{equation} 
{\cal E}_k C_k = C_{k-1}+C_{k+1} 
\label{effLloyd} 
\end{equation} 
It is a  Lloyd model with a $k$-dependent distribution of on-site 
energies characterized by parameter 
\begin{equation} 
\delta_k(E)=\frac{\delta(E)}{V} \frac{n(k+1)}{ d(k)} 
\label{Deltak} 
\end{equation} 
which in the limit $k\rightarrow\infty$ has the property 
\begin{equation} 
\delta_k(E) \sim  
\sqrt{\frac{N}{6k}}\frac{\Delta}{V} 
\label{Deltaklim} 
\end{equation} 
 
We generalize the solution of the Lloyd model \cite{ishii}  to the present 
case of site-dependent distribution of energies.  We observe that the distribution 
parameter depends weakly on the site number $k$ and use 
the site-dependent localization length ansatz: 
\begin{equation} 
1/\xi_k = \ln\frac{|C_k|}{|C_{k+1}|} = 
\cosh^{-1}\sqrt{1+\frac{\delta_k^2}{4}} \approx 
\ln\delta_k=\ln\sqrt{\frac{N}{6k}}\frac{\Delta}{V} 
 \label{locl} 
\end{equation} 
The last approximation is justified in the weak-coupling  limit $V/\Delta\ll 1$. 
 
The validity of the site-dependent localization length assumption is 
checked numerically. The on-site 
energies ${\cal E}_k$ were generated from the Cauchy 
distribution (\ref{Cauchy}) with parameter $\delta_k$ given by 
(\ref{Deltaklim}). The inverse localization length  $1/\xi$ 
(Lyapunov exponent) 
was calculated using the standard technique of transfer matrices \cite{ishii} 
and it was 
found that the site-dependent ansatz (\ref{locl}) 
describes well (at least qualitatively) the behavior of the 
localization length.
Thus for independently distributed energies the 
typical behavior of the wave functions 
is described by  coefficients that decay as 
\begin{equation} 
C_k\sim  
e^{-k/\xi_k}=\exp\left({-k\log\sqrt{\frac{N}{6k}}\frac{\Delta}{V}}\right). 
\label{coeffk} 
\end{equation} 
In the strong coupling limit similar considerations lead to
\begin{equation} 
C_k\sim 
e^{-k/\xi_k}=\exp\left({-k\sqrt{\frac{N}{24k}}\frac{\Delta}{V}}\right)\;,
\label{coeffkstrong} 
\end{equation}
but in this limit these require some further justification. 
The importance of this result is that numerical studies of more 
realistic many-particle energy distributions confirm the validity of 
(\ref{coeffk}).  This will be the subject of the next section. 
 
\section{Effects of energy correlations} 
 
 In the last section we considered  a model in which the diagonal matrix elements 
$E_\prt{\lambda}$ are independent random variables. 
We did not specify the distribution law of these energies, since 
the solution does not depend on fine details, but rather on 
general properties of the distribution. We would like to stress 
that the assumption of statistical independence of the 
energies $E_\prt{\lambda}$ is not realistic for a many-body 
system. To illustrate this statement   a simple argument can be 
given. Suppose we are dealing with a disordered mesoscopic system 
and the one-particle spectrum is random, i.e. consists of levels 
with independently distributed spacings $\Delta_j=E_j-E_{j-1}$. 
The generalization of the expression (\ref{Epart}) for the  energy 
of the partition $\prt{\lambda}$ for this case is 
\begin{equation}\label{Erand} 
E_\prt{\lambda}=\sum_i \left(E_{1-i+\lambda_i}-E_{1-i}\right)= 
\sum_i\sum_{j=2-i}^{\lambda_i+1-i}\Delta_j 
\end{equation} 
In order to describe the energies $E_\prt{\lambda}$ in the 
subspace $N$ we need $2N$ random quantities $\Delta_j$. On the 
other hand the number of states in this subspace is given by 
$p(N)$, the number of partitions of $N$ which grows exponentially 
for $N\rightarrow\infty$. Therefore there exist much more 
many-body energies $E_\prt{\lambda}$ than independent level spacings. 
Therefore  the 
many-particle energies $E_\prt{\lambda}$, which are the sums of the 
one-particle energies, are statistically dependent. 
 
The analytical calculation of the (joint)  distribution of the 
parameters ${\cal E}_k$ defined in (\ref{Ek}) for the effective 
tight-binding model (\ref{effSchr}) seems to be hopeless. 
Nevertheless we expect some features of this  distribution, 
based on the formula (\ref{Ek}). For an eigenstate the eigenvalue 
$E$ is in the interval 
$[-1+E_\prt{\lambda}^{min},+1+E_\prt{\lambda}^{max}]$, 
where $E_\prt{\lambda}^{min}$ and $E_\prt{\lambda}^{max}$ are the 
minimal and maximal values of $E_\prt{\lambda}$. The value of $E$ 
can be very close one of the $E_\prt{\lambda}$. Therefore the 
terms in the sum on the RHS of (\ref{Ek}) can be positive or 
negative and they strongly fluctuate in magnitude. Therefore 
there is a finite probability that this sum will take a value in 
the vicinity of zero. Consequently it is expected that 
the distribution $P({\cal E}_k)$ is  broad  with 
a diverging second moment or variance: 
\begin{equation}\label{variance} 
  \Delta {\cal E}_k^2=\<({\cal E}_k-\<{\cal E}_k\>)^2\>\rightarrow\infty 
\end{equation} 
 
Another important issue is the effective statistical independence 
of ${\cal E}_k$ for different $k$-s. It is   known  for Anderson 
localization that the long-range correlations between 
the on-site energies change dramatically the criteria for the 
onset of the localization (see \cite{izrailev} and references therein). 
The measure of statistical dependence of energies in our case  is the 
correlation matrix defined as 
\begin{equation}\label{correlation} 
C_{kl} =\frac{\<({\cal E}_k-\<{\cal E}_k\>) ({\cal E}_l-\<{\cal 
E}_l\>)\>} { \Delta {\cal E}_k \Delta {\cal E}_l} 
\end{equation} 
Because of the strong fluctuations between the terms of the sum on 
the RHS of (\ref{Ek}), that are different for the various groups, 
it is reasonable that the off-diagonal terms of the correlation 
function are much smaller than the diagonal ones. Usually for 
localization only pair correlations are important \cite{naama}.
In the view of the diverging variance we assume that $C_{kl}=\delta_{kl}$, an 
assumption that will be tested numerically. 
 
\begin{figure} 
  \centerline{ 
              \psfig{figure=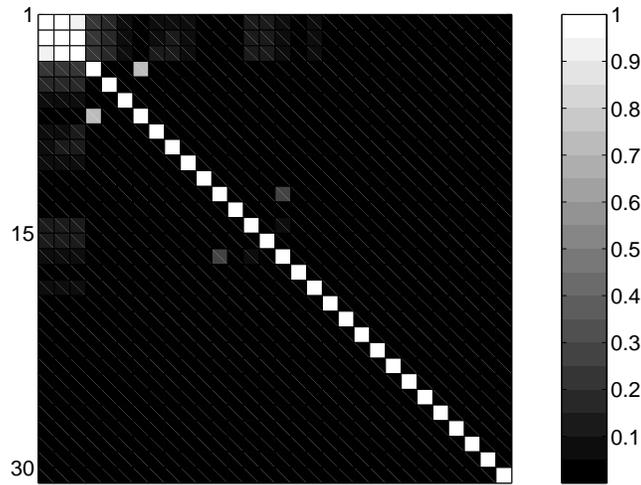   ,width=8.6cm    }
             } 
\hspace{10pt} 
  \caption{The absolute value of correlation matrix elements $C_{kl}$ defined in 
  (\ref{correlation}) of effective energies ${\cal E}_k$ for $N=30$. } 
\label{correl} 
\end{figure}

\begin{figure} 
  \centerline{ 
              \psfig{figure=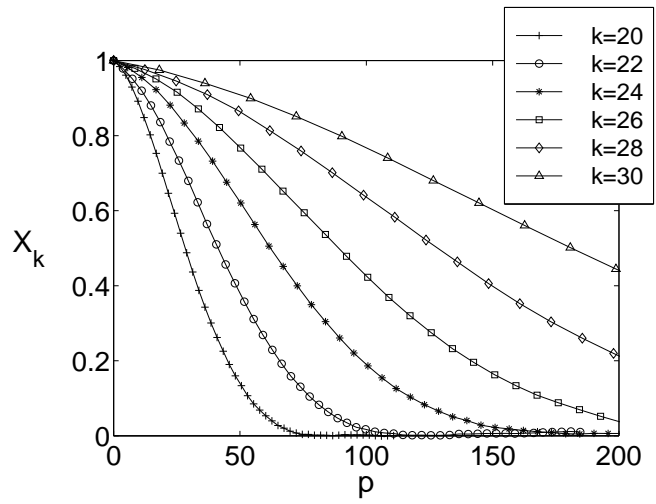   ,width=8.6cm    } 
             } 
\hspace{10pt} 
  \caption{Characteristic function $X_k (p)$ of the 
  effective energy distribution for some values of $k$. } 
\label{charact} 
\end{figure}

These facts were  checked numerically and in the 
approximation of independent energies the effective distribution 
$P({\cal E}_k)$ can be calculated. For this purpose $M=1000$ 
realizations of $N=30$ level spacings $\Delta_j$ were generated 
from the exponential distribution: 
\begin{equation} 
P(\Delta)=e^{-\Delta} 
\label{Exp} 
\end{equation} 
with $\bar{\Delta}=1$. Then we generated numerically 
 the whole list of $p(30)=5604$ partitions of $N=30$ and 
for each realization of level spacings the energies 
$E_\prt{\lambda}$ were calculated using (\ref{Erand}). Out of 
these energies the effective energies ${\cal E }_k$ were obtained 
using the definition (\ref{Ek}) for $E=0$. Several statistical 
tests were performed and results are shown in 
Fig.~\ref{correl}-\ref{charact}. 
 
The correlation matrix of the energies  ${\cal E}_k$ was computed. 
The $\<\ldots \>$  averages were performed over realizations of 
the level spacings. The result is shown in Fig.~\ref{correl}. It 
is plausible that the correlations are of short range in the space 
of indices $k$ and  only the diagonal elements of $C_{kl}$ are of 
appreciable magnitude. This behavior can be attributed to the 
diverging variance.

\begin{figure} 
  \centerline{ 
              \psfig{figure=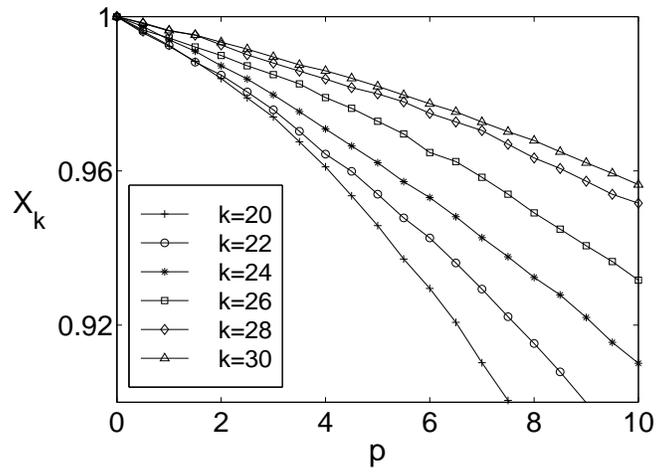   ,width=8.6cm    } 
             } 
\hspace{10pt} 
  \caption{The same as on the previous figure for a smaller range of $p$. } 
\label{charactsmall} 
\end{figure}

Due to the  fact that the distribution of the energies ${\cal E}_k$ is expected 
to have a  diverging variance the distribution $P({\cal E}_k)$ 
cannot be calculated from a histogram. 
Instead its characteristic function 
\begin{equation}\label{character} 
  X_k (p)=\<e^{ip {\cal E}_k}\> 
\end{equation} 
was computed and plotted for several values of $k$ in 
Fig.~\ref{charact} as a function of $p$. The diverging second 
moment due to the fact that $P(x)\propto 1/x^2$ manifests itself in 
the discontinuity of the derivative at 
$p=0$ as can be seen from the Fig~\ref{charactsmall}. Therefore we 
have strong evidence that the energies ${\cal E}_k$ have a broad 
distribution.

\begin{figure} 
  \centerline{ 
              \psfig{figure=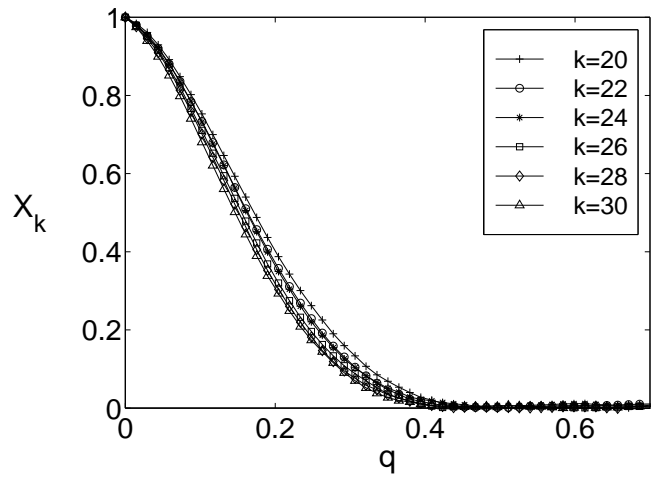   ,width=8.6cm    } 
             } 
\hspace{10pt} 
  \caption{The characteristic function $X_k (p)$ as a function of the scaled variable 
    $q=p\Delta /n(k) $ for values of $k$ used in  Figs~\ref{charact}-\ref{charactsmall}} 
\label{scaling} 
\end{figure}

Observing  the behavior of the characteristic function shown on 
the \fig{charact}  and \fig{charactsmall} we introduced  a scaling 
ansatz for the distribution function: 
\begin{equation}\label{scalingP} 
  P({\cal E}_k)=P_k({\cal E}) =\frac{n(k)}{\Delta}\widetilde{P} 
  \left( n(k)\frac{\cal E}{\Delta}\right). 
\end{equation} 
The universal function 
$\widetilde{P}(x)$ is independent of $k$. The scaling of the 
distribution implies that the characteristic function 
(\ref{character}) satisfies 
\begin{equation}\label{scalingX} 
  X_k (p) = \widetilde{X} \left(\frac{p\Delta }{n(k)}\right),\;\;\;\;\; 
  \widetilde{X} (q)=\int_{-\infty}^\infty dx\, \widetilde{P}(x) 
  e^{iqx} 
\end{equation} 
which is easy to check by plotting $X_k (p)$ as a function of 
$q=p\Delta/n(k)$ for different values of $k$. These curves are 
displayed in Fig.~\ref{scaling} and are indeed found to be close 
to each other in spite of  the fact that $n(k)$ ranges from $n(1)=1$ to $n(30)=1039$.

We found numerically that 
the universal characteristic function can be well approximated by 
the general form 
\begin{equation}\label{formX} 
  \widetilde{X} (q) = \exp\left(-a|q|-\half{b q^2}\right) 
\end{equation} 
The value of parameters were found to be $a\approx 1.15$, $b\approx 
6.32$. This is the  characteristic function of a distribution of a 
sum of two random variables $x_1$ and $x_2$, where $x_1$ is drawn from 
Cauchy distribution with probability density 
\begin{equation} 
p_1(x)=\frac{1}{\pi}\frac{a}{x^2+a^2} 
\label{normal} 
\end{equation} 
and the second is distributed normally according to 
\begin{equation} 
p_2(x) =\frac{1}{\sqrt{2\pi b}}e^{-\frac{x^2}{2b}}. 
\label{cauchy} 
\end{equation} 
The distribution of $x=x_1+x_2$ is thus given by 
\begin{equation} 
\tilde{P}(x)=\int_{-\infty}^{+\infty} dy p_1(x-y)p_2(y) 
\label{formP} 
\end{equation} 
and its characteristic function is (\ref{formX}). It 
 has a diverging variance due to the fact that $\tilde{P}(x)$ 
behaves as $\frac{1}{\pi}\frac{a}{x^2}$ for large $|x|$.

The scaling factor $\Delta/n(k)$ and  the limiting behavior of 
the distribution function are  the same as those found in the case 
of mutually independent energies. This allows us to conclude that 
the behavior of the coefficients $C_k$ given by (\ref{coeffk}), 
that was obtained in  the model with uncorrelated 
$E_\prt{\lambda}$, is satisfied also by the eigenfunctions of the 
more realistic model of the present section. 
 
\section{Discussion} 
 
The localization properties of the eigenstates of the modified 
interacting chiral Tomonaga-Luttinger model were studied. For 
total energy $N\Delta$ subspace we defined an effective lattice of 
Slater determinants labeled by partitions of $N$ which 
represent the eigenstates of the system in the absence of 
interactions. We have observed that in the case of linear 
dispersion the eigenstates of the interacting model described by 
the bosons are extended over this lattice. When the random 
one-particle levels are introduced instead of the equidistant spectrum, 
the bosons are no more the eigenstates and the problem of 
diagonalizing of the Hamiltonian. In order to 
achieve this goal we have grouped the different sites (partitions) 
according to the number of ones, representing the distance of the 
most excited particle from the Fermi surface in the original 
Slater determinant. This is to be contrasted with the grouping of 
many-body states into generations according to the number of 
excited particles and holes as was firstly done in \cite{altsh}. 
Each resulting group of partitions was given a label $k$. The 
interactions were shown to be of short range in $k$. The 
interactions were further simplified by assuming that they 
interconnect only groups of partitions corresponding to the 
adjacent values of the index $k$ (and the interactions between 
groups $k$ and $k\pm 2$ were neglected). Assuming further that 
the states are extended within each group $k$ we obtained an 
effective tight-binding Anderson model where each group $k$ has an 
effective random energy and the hopping takes place between the 
nearest  neighbors only. In the assumption of independently 
distributed on-site energies the model is found to be close to the 
Lloyd model characterized by a broad distribution of the on-site 
energies. The main feature of our effective model is that  the 
distribution of energies is $k$-dependent.  We employed the 
site-dependent localization length assumption, which consists in 
assuming the parameters of  distribution  change slowly, so 
locally the localization length is given by the same formula as 
for the Lloyd model with parameters of the local distribution 
instead of the uniform global one. Up to an unknown numerical factor 
this solution describes well the behavior of the localization 
length as a function of $k$ and it enabled us to show that the 
eigenstates are localized in the space of groups of partitions, 
$k$. The weak-coupling and the strong-coupling regimes can be 
identified according to the  value of the parameter $V/\Delta$. In 
the weak-coupling regime $V\ll\Delta$, the amplitudes decay as 
$\exp\left({-k\log\sqrt{\frac{N}{6k}}\frac{\Delta}{V}}\right)$, 
namely nearly exponentially. In the strong coupling regime 
$V\gg\Delta$, this behavior changes to 
$\exp\left({-\sqrt{\frac{Nk}{24}}\frac{\Delta}{V}}\right)$. 
 
One important point which was discussed is the effect of 
correlations  between the many-body energies of the 
non-interacting system. In the studies devoted to the localization 
in many-body systems the effective on-site energies corresponding 
to the eigenstates of the non-interacting system are assumed to be 
independent random variables, so  their spectrum is described by a 
Poissonian sequence. This is not the case for the true many-body 
fermionic system, since these energies are sums of independent 
energies of one-particle excitations. The number of independent 
random variables is therefore much less than the number of 
different many-body states. This situation is similar in some 
sense to  one faced in random walk problems, where the increments 
are distributed independently, while there is a strong correlation 
between the positions of the random walk at different times. 
In our work we considered this problem numerically and 
found that although the correlation between the exact energies 
corresponding to the eigenstates of the free model are not {\em a 
priori} negligible the random energies of the effective sites $k$ 
are found to be very weakly correlated. Moreover it was found that 
the parameters of the distribution of the effective energy of the 
group $k$ scale with the number of the partitions inside the group 
in the same way as they do in the case of independent many-body 
energies. This led us to the conclusion that our result for the 
localization of the eigenstates are valid for more realistic 
assumptions on the distribution of the many-body energies. 
 
The main open problem arising from the present discussion is the 
study of the interacting electrons in dimensions higher than one 
by the methods of high-dimensional bosonization. In this technique 
\cite{Castro,Hough1,Hough2,Schoen1,Schoen2} the fermionic system 
is described as an infinite set of one-dimensional Luttinger 
models attached to a point on the Fermi surface. Therefore the 
interactions mix the states belonging to different points in 
addition to the coupling within the same point (represented by a 
Luttinger model), which can be treated within the framework 
developed in the present paper. This possibly can be a mechanism 
for the delocalization of the states and  a resulting localization 
transition must be compared with the transition encountered in the 
recent studies \cite{altsh,mirl,berk,pascaud} of interacting 
electrons in finite systems. 

Another question to be explored in the future is the nature of 
localization in the Hilbert space for single-particle dispersions 
that are more realistic than the random ones.
 
\section*{Acknowledgements}
The authors would like to thank Alex Kamenev for useful remarks
and Eric Akkermans for stimulating discussions in the initial stages 
of the project. This research was supported in part by the
U.S.--Israel Binational Science Foundation (BSF), by the Minerva
Center for Non-linear Physics of Complex Systems, by the Israel
Science Foundation, by the Niedersachsen Ministry of Science
(Germany) and by the Fund for Promotion of Research at the
Technion.

\appendix
\section{A property for characters of $S_N$} 
 
In this appendix we show that the property (\ref{theproperty}) 
\begin{eqnarray*}\sum_{i,j}\eta_\prt{\ldots\lambda_i-m\ldots} 
\eta_\prt{\ldots\lambda_i-m\ldots\lambda_j+m\ldots} 
\chi^\prt{\ldots\lambda_i-m\ldots\lambda_j+m\ldots}_\prt{l}= 
 mn_m \chi^\prt{\lambda}_\prt{l} 
\end{eqnarray*} 
is satisfied for the characters of the symmetric group $S_N$. 
We did not find this property  in any  textbook on 
the representations of groups, but its validity follows from the known 
properties of the 
characters of $S_N$ found, for example, in \cite{littlewood} and \cite{hammermesh}. 
We  present the proof of this property here for the sake of 
completeness of the discussion. 
 
The Frobenius formula as it is written in the classic textbooks on group 
theory \cite{hammermesh,littlewood} states the following identity between two 
antisymmetric functions of $N$-dimensional vector $z=(z_1,z_2,\ldots,z_N)$: 
\begin{equation} 
S_\prt{l} (z) D(z)=\sum_\prt{\lambda}\chi^\prt{\lambda}_\prt{l} \Psi_\prt{\lambda} (z) 
\label{theFrobenius} 
\end{equation} 
where $\chi_\prt{l}^\prt{\lambda}$ is the character of the class 
$\prt{l}$ of the symmetric group $S_N$ in the  irreducible 
representation $\prt{\lambda}$.  With the partition 
$\prt{\lambda}$ one associates the totally antisymmetric function: 
\begin{equation} 
\Psi_\prt{\lambda} (z)\equiv \sum_P \mbox{sgn}(P)\, 
z_{P(1)}^{\lambda_1+N-1}z_{P(2)}^{\lambda_2+N-2} \ldots 
z_{P(N)}^{\lambda_N} \label{sldet} 
\end{equation} 
where the sum runs over all the permutations of $1,2,\ldots,N$, 
the sign of permutation being $\pm$ for an even/odd permutation. 
This is, up to a normalization constant, a Slater determinant 
constructed out of one-particle wave-functions 
$z^{n_i}=z^{\lambda_i+N-i}$, where $z=e^{i\theta}$. The factor $D(z)$ in 
(\ref{theFrobenius}) is the Vandermonde 
determinant: 
\begin{equation}\label{VdM} 
  D(z)=\left|\begin{array}{ccccc} 
    1      & 1     & \cdots & 1 \\ 
    z_1    & z_2   & \cdots & z_N \\ 
    z_1^2  & z_2^2 & \cdots & z_N^2 \\ 
    \vdots & \vdots  & \ddots & \vdots \\ 
    z_1^{N-1} & z_2^{N-1} & \cdots & z_N^{N-1} \ 
  \end{array}\right|=\prod_{i<j}(z_i - z_j) 
\end{equation} 
representing the ground state of $N$ fermions. Let $S_m $ be a 
power sum of variables $z_j$ defined as follows: 
\begin{equation}\label{sm} 
  S_m (z) = \sum_j z_j^m 
\end{equation} 
The function $S_m(z)$ is totally symmetric function of the  $z_j$-s. 
These functions are linearly independent as can be checked by 
calculating the Jacobian: 
\begin{equation}\label{jacob} 
  \left|\frac{\partial S_m}{\partial z_j}\right|=N!D(z) . 
\end{equation} 
The class $\prt{l}$ is characterized by 
$n_1,n_2,\ldots,n_m,\ldots$ cycles of length 
$1,2,\ldots,m,\ldots$. For this class  the totally 
symmetric function $S_\prt{l}$ is defined as a product of power 
sums 
\begin{equation} 
S_\prt{l} (z) = S_1^{n_1} S_2^{n_2}\ldots 
S_N^{n_N}=\prod_{m=1}^\infty S_m^{n_m} 
\label{sl} 
\end{equation} 
Since for different $m$-s the power sums $S-m$ are linearly 
independent so are the monomials $S_\prt{l}$ for different 
partitions $\prt{l}$. 
 
The Frobenius formula (\ref{theFrobenius}) can be inverted with 
help of the orthogonality relation of the characters: 
\begin{equation} 
\Psi_\prt{\lambda} (z)= 
\frac{1}{g}\sum_\prt{l}g_\prt{l}\chi^\prt{\lambda}_\prt{l}S_\prt{l}(z) 
D(z) \label{invFrobenius} 
\end{equation} 
where $g=N!$ is the order of the $S_N$ and $g_\prt{l}$ is the 
number of elements of $S_N$ in the class $\prt{l}$ given by 
(\ref{gl}).

We turn now to develop a recurrence relation between the 
characters. For this purpose it is instructive to consider the 
product of the antisymmetric function $\Psi_\prt{\mu} (z)$ and a 
power sum $S_m(z)$. Since the power sum is totally symmetric the 
summands can be rearranged in the order set by each permutation 
and one obtains 
\begin{eqnarray} 
S_m(z) \Psi_\prt{\mu} (z) &=& \sum_P \mbox{sgn}(P)  \left( \sum_i 
z_{P(i)}^m\right)  z_{P(1)}^{\mu_1+N-1} z_{P(2)}^{\mu_2+N-2} 
\ldots z_{P(N)}^{\mu_N} = \nonumber \\ &=& \sum_i 
\eta_\prt{\ldots\mu_i+m\ldots} \Psi_\prt{\ldots 
\mu_i+m\ldots} (z) \label{thefact} 
\end{eqnarray} 
where $\eta_\prt{\ldots\mu_i+m\ldots}=-1,0$ or $+1$ is determined 
accordingly to the rules of Section 2.

Having established the relation between the Slater determinants 
$\Psi_\prt{\mu}$ for various partitions we turn to develop a 
recurrence relation between the characters. Consider class 
$\prt{l}$ with at least one $m$-cycle. Let us denote by 
$\prt{l_m'}$ the class obtained by removing one $m$-cycle from 
$\prt{l}$. It is then clear that 
\begin{equation}\label{slprime} 
  S_\prt{l}=S_m S_\prt{l_m'}= 
  \left(\sum_j z_j^m\right)S_\prt{l_m'} 
\end{equation} 
The partition $\prt{l_m'}$ defines a class in the symmetry group 
$S_{N-m}$. The Frobenius formula (\ref{theFrobenius}) for this 
class reads 
\begin{equation}\label{theFrobenius'} 
  S_\prt{l_m'} D(z) = \sum_\prt{\mu'}\chi_\prt{l_m'}^\prt{\mu'} 
  \Psi_\prt{\mu'} 
\end{equation} 
while for  $\prt{l}$ it can be rewritten with the help of 
(\ref{thefact}) in the form 
\begin{eqnarray} 
  S_\prt{l} D(z) &=& S_m S_\prt{l_m'} D(z) =\sum_\prt{\mu'}\chi_\prt{l_m'}^\prt{\mu'} 
  S_m  \Psi_\prt{\mu'} = 
  \nonumber\\ 
  &=& \sum_\prt{\mu'}\chi_\prt{l_m'}^\prt{\mu'} 
  \sum_i\eta_\prt{\ldots\mu'_i+m\ldots} \Psi_\prt{\ldots 
  \mu'_i+m\ldots} 
  \label{theform} 
\end{eqnarray} 
Comparing the coefficients of the $\Psi_\prt{\mu'}$ in 
(\ref{theform}) with those in (\ref{theFrobenius}) we find the 
recurrence relation of the characters 
\begin{equation} 
\chi^\prt{\lambda}_\prt{l} = \sum_i 
\eta_\prt{\ldots\lambda_i-m\ldots} 
\chi^\prt{\ldots\lambda_i-m\ldots}_\prt{l'_m} \label{recurrence} 
\end{equation} 
This relation yields the characters of $S_N$ in terms of the 
characters of the symmetric group $S_{N-m}$. 
 
In order to derive (\ref{theproperty})  take some particular $m$ 
and let each partition $\prt{l}$ contain $m$ as its part 
$n^\prt{l}_m$ times. For each $\prt{\lambda}$ consider the 
following sum over partitions: 
\begin{equation} 
F_\prt{\lambda}(z) = 
\frac{1}{g}\sum_\prt{l}g_\prt{l} m n^\prt{l}_m 
\chi^\prt{\lambda}_\prt{l}S_\prt{l}(z) 
\label{fz} 
\end{equation} 
If $\prt{l}$ has no parts equal to $m$, $n^\prt{l}_m=0$ and 
$\prt{l}$  does not contribute to this sum. We want to factor out 
$S_m (z)$ from the sum and for this purpose we can write 
$S_\prt{l} (z) = S_m (z) S_\prt{l_m'} (z)$, so that the sum is 
over $\prt{l_m'}$. With the help of (\ref{gl}) we see that the 
expansion coefficients satisfy 
\begin{eqnarray} 
&&g_\prt{l} m n_m /g =m n_m\prod_{k=1}^{\infty} \frac{1}{k^{n_k} 
n_k!} = \frac{m n_m}{m^{n_m} n_m!}\prod_{k\neq m} \frac{1}{k^{n_k} 
n_k!} \nonumber\\ &&= \frac{1}{m^{n_m-1} 
\left(n_m-1\right)!}\prod_{k\neq m} \frac{1}{k^{n_k} 
n_k!}=g_\prt{l_m'}/g' \label{glprime} 
\end{eqnarray} 
where $g'=(N-m)!$. The sum in (\ref{fz}) takes the form 
\begin{equation} 
F_\prt{\lambda}(z) = \left(\sum_j z_j^m\right) 
\frac{1}{g'}\sum_\prt{l_m'}g_\prt{l_m'} 
\chi^\prt{\lambda}_\prt{l}S_\prt{l_m'}(z) \label{fz1} 
\end{equation} 
Application of the recurrence relation (\ref{recurrence}) to the 
characters in the sum together with the inverse Frobenius formula 
(\ref{invFrobenius}) yields 
\begin{eqnarray} 
F_\prt{\lambda}(z) D(z) &=& S_m (z) \sum_i 
\eta_\prt{\ldots\lambda_i-m\ldots} 
\frac{1}{g'}\sum_\prt{l_m'}g_\prt{l_m'} 
\chi^\prt{\ldots\lambda_i-m\ldots}_\prt{l_m'}S_\prt{l_m'} (z) 
D(z)= 
 \nonumber\\ &=& \sum_i \eta_\prt{\ldots\lambda_i-m\ldots} 
 S_m (z) \Psi_\prt{\ldots\lambda_i-m\ldots}(z) 
 \label{fz2} 
\end{eqnarray} 
The relation (\ref{thefact}) is applied to show that 
\begin{equation} 
F_\prt{\lambda}(z) D(z) = 
\sum_{i,j}\eta_\prt{\ldots\lambda_i-m\ldots} 
\eta_\prt{\ldots\lambda_i-m\ldots\lambda_j+m\ldots} 
\Psi_\prt{\ldots\lambda_i-m\ldots\lambda_j+m\ldots} (z) 
\label{fz3} 
\end{equation} 
Application of the inverse Frobenius formula (\ref{invFrobenius}) yields 
\begin{equation} 
  F_\prt{\lambda}(z) 
  =\frac{1}{g}\sum_{ij} \sum_\prt{l}\eta_\prt{\ldots\lambda_i-m\ldots} 
  \eta_\prt{\ldots\lambda_i-m\ldots\lambda_j+m\ldots} 
   g_\prt{l}\chi_\prt{l}^\prt{\ldots\lambda_i-m\ldots\lambda_j+m\ldots} 
  S_\prt{l} (z) 
  \label{fz4} 
\end{equation} 
Making use of the linear independence of $S_\prt{l}$ for different 
partitions $\prt{l}$ and equating the coefficients of the 
$S_\prt{l} (z)$ with the corresponding coefficients in (\ref{fz}) 
results in the required relation (\ref{theproperty}).

\section{Number of distinct parts of partitions.} 
 
In this appendix we show that $d(N)$ -- the number of distinct 
parts in the partitions of $N$ with parts $\lambda_i\geq 2$ equals 
$p(N-2)$ -- the number of unrestricted partitions of $N-2$. 
Together with the generating function (\ref{Gen}) for the number 
of partitions let us introduce the generating function 
\begin{equation} 
\tilde{p}(z,q) = \sum_{N=1}^{\infty}\sum_{d=1}^{\infty} 
\tilde{p}(N,d) q^N z^d \label{Gen1} 
\end{equation} 
for the number $\tilde{p}(N,d)$ of partitions of $N$ having 
entries that satisfy $\lambda_i\geq 2$ and  having exactly $d$ 
distinct parts. For example for $N=6$ and $d=2$ this number is 
$\tilde{p}(N,d)=1$, corresponding to the partition $(4,2)$ as can 
be seen from the list (\ref{Partitions}). The number of distinct 
parts $d(N)$ in all partitions of $N$ with parts $\ge 2$ is then 
given by 
\begin{equation} 
d(N) = \sum_{d=1}^{N} d \tilde{p}(N,d) \label{dn} 
\end{equation} 
It is generated by the following function 
\begin{equation} 
d(q) = \sum_{N=1}^{\infty} d(N) q^N = \left. z\frac{d }{d z} 
\tilde{p}(z,q) \right|_{z=1} \label{dq} 
\end{equation} 
The explicit expression for $\tilde{p}(z,q)$ can be found easily summing 
directly over all the appropriate partitions: 
\begin{equation} 
\tilde{p}(z,q) = \sum_{\prt{\lambda},\lambda_i\ge 2} 
q^{|\prt{\lambda}|} z^{d\prt{\lambda}} \label{Gen2} 
\end{equation} 
where $d\prt{\lambda}$ is the number of distinct parts in 
$\prt{\lambda}$ and $|\prt{\lambda}|=\sum_i \lambda_i$. It is 
convenient to use  the sets of  ``occupation numbers'' $\{n_m\}$, 
where $n_m$ -- the number of times $m$ appears in the partition 
$\prt{\lambda}$. Noting that $|\prt{\lambda}|=\sum_i \lambda_i= 
\sum_m m n_m$ and $d\prt{\lambda}=\sum_m \theta (n_m)$, where 
$\theta(x)$ is the Heavyside step function we rewrite (\ref{Gen2}) 
as a sum over all the configurations $\{n_m\}$, with $m\geq 2$ 
\begin{eqnarray} 
\tilde{p}(z,q) &=& \sum_{\{n_m\}, m \ge 2} q^{\sum_{m=2}^\infty m 
n_m} z^{\sum_{m=2}^\infty \theta(n_m)} = \prod_{m=2}^\infty 
\sum_{n_m=0}^\infty q^{m n_m} z^{\theta(n_m)}=\nonumber \\ &=& 
\prod_{m=2}^\infty \left(1+\frac{zq^m}{1-q^m}\right) \label{Gen3} 
\end{eqnarray} 
Differentiating this expression with respect to $z$ at $z=1$ we 
get 
\begin{equation} 
d(q) =\left. z\frac{d }{d z} \prod_{m=2}^\infty 
\left(1+\frac{zq^m}{1-q^m}\right) \right|_{z=1}=q^2 
\prod_{m=1}^\infty \frac{1}{1-q^m}=q^2 p(q) \label{dq1} 
\end{equation} 
 where $p(q)$ is given by (\ref{Gen}). The function 
$q^2 p(q)$ is  the generating function for $p(N-2)$. The equality 
of the generating functions $d(q)$ and $q^2 p(q)$ implies the equality of the 
coefficients: $d(N) = p(N-2)$, i.e. the number of different parts 
in all partitions of $N$ with entries $\ge 2$ is given by the 
number of unrestricted partitions of $N-2$. Applying this 
argumentation to all partitions ending with $N-k$ ones yields 
$d(k)=p(k-2)$.

\section{Distribution of harmonic sum of independent random variables.} 
 
We consider the harmonic mean $X$ defined as 
\begin{equation} 
\frac{1}{X}=\frac{1}{N}\sum_{n=1}^N\frac{1}{x_n} 
\label{harmsum} 
\end{equation} 
of $N$ independent random variables  $x_1, x_2,\ldots x_N$  drawn from the 
common distribution with probability density $P(x)$. The only requirement 
we impose on $P(x)$ is that 
\begin{equation} 
\lim_{x\rightarrow 0} P(x)=C > 0 
\label{limp} 
\end{equation} 
Consider the random variable $y=1/x$. 
Its distribution function $Q(y)$ can be calculated by the standard 
methods. For 
$y\rightarrow\pm\infty$ it  has the  asymptotic behavior 
\begin{equation} 
Q(y)\rightarrow\frac{ C}{y^2} 
\label{assympt} 
\end{equation} 
and  its second moment diverges. The general theory of broad 
distributions \cite{bouchaud,levi,gnedenko} predicts that in the limit 
$N\rightarrow\infty$ the probability density of the sum (\ref{harmsum}) $Y=1/X$ is 
Lorenzian 
\begin{equation} 
Q_N (Y)  = \frac{1}{\pi}\frac{\pi C}{(\pi C)^2+Y^2} 
\label{lorenzian} 
\end{equation} 
and consequently the probability density $P_N (X))$ of variable 
$X=1/Y$ is again a Lorenzian: 
\begin{equation} 
P_N (X)  = \frac{1}{\pi}\frac{\delta}{\delta^2+X^2} 
\label{lorenziaN} 
\end{equation} 
characterized by the  parameter $\delta$ related to the original 
distribution $P(x)$ by the formula: 
\begin{equation}\label{delta} 
  \delta=\frac{1}{\pi C} 
\end{equation} 
In particular for the uniform distribution  (\ref{PE}) it takes 
the value $\delta=\Delta/\pi$.

\newpage
\newpage
\newpage
\end{document}